# DFT exploration of novel direct band gap semiconducting halide double perovskites, $A_2AgIrCl_6$ (A = Cs, Rb, K), for solar cells application

M. A. Rayhan[1,2,3], M. M. Hossain[1,2], M. M. Uddin[1,2], S. H. Naqib[2,4], M. A Ali[1,2,*]

[1]Department of Physics, Chittagong University of Engineering and Technology (CUET), Chattogram-4349, Bangladesh
[2]Advanced Computational Materials Research Laboratory (ACMRL), Department of Physics, Chittagong University of Engineering and Technology (CUET), Chattogram-4349, Bangladesh
[3]Department of Arts & Sciences, Bangladesh Army University of Science and Technology (BAUST) Saidpur Cantonment, Saidpur-5310, Nilphamari
[4]Department of Physics, University of Rajshahi, Rajshahi-6205, Bangladesh

**Abstract**

Double perovskite halides are promising materials for renewable energy production, meeting the criteria to address energy scarcity issues. As a result, studying these halides could be useful for optoelectronic and solar cell applications. In this study, we investigated the structural, mechanical, thermodynamic, electronic, and optical properties of $A_2AgIrCl_6$ (A = Cs, Rb, K) double perovskite halides using density functional theory calculations with the full-potential linearized augmented plane-wave (FP-LAPW) approach, aiming to evaluate their suitability for renewable energy devices. The Goldsmith tolerance factor, octahedral factor, and new tolerance factor have confirmed the cubic stability of the predicted compounds. We have also verified the thermodynamic stability of these compounds by calculating the formation enthalpy, binding energy, and phonon dispersion curves. Additionally, Born-Huang stability requirements on stiffness constants confirmed the mechanical stability of the titled compounds. To predict the accurate optoelectronic properties, we employed the TB-mBJ potential. The electronic band structure calculations revealed that the titled halides exhibit a direct band gap semiconducting nature with values of 1.43 eV, 1.50 eV, and 1.55 eV for $Cs_2AgIrCl_6$, $Rb_2AgIrCl_6$, and $K_2AgIrCl_6$, respectively. Besides, all these compounds showed remarkably low effective electron masses, indicating their potential for high carrier mobility. Furthermore, the optical properties of $A_2AgIrCl_6$ (A = Cs, Rb, K) compounds demonstrated very low reflectivity and excellent light absorption coefficients ($10^5$ cm$^{-1}$) in the visible light spectrum, suggesting their suitability as an absorbing layer in solar cells. The photoconductivity and absorption spectra of these compounds validate the accuracy of our band structure results.

## 1. Introduction

Earth's limited energy supplies are stressed by the world's increasing energy consumption caused by population expansion and growing living standards [1,2]. Over the previous 150 years, a notable increase in energy use has outpaced population growth. Fossil fuels, mainly gas and oil, account for over 80% of global use, which has issues with supply and impact on the environment. This clarifies how urgently renewable energy sources like geothermal, wind, and solar power need to be used [3–5]. Conservation and the use of renewable energy are becoming increasingly important to both governments and citizens. Of the possible substitutes, metal halide perovskites have garnered substantial interest and have been thoroughly studied in the past several years [6]. These materials have drawn much attention over the past ten years due to their unique characteristics, adaptable synthesis methods, and configurable device designs. Perovskites based on lead have become a viable and affordable alternative for high-efficiency solar cells. Meanwhile, their commercialization has been hampered by problems with lead toxicity and chemical instability despite their excellent efficiency [7–9]. Even though a number of lead-free perovskites have been suggested as a solution to the toxicity issue, they still have difficulty reaching comparable levels of efficiency.

Trivalent cations like $Bi^{3+}$ and $Sb^{3+}$ have recently been used to create 2D layered lead-free halide perovskites[10]. Researchers have also incorporated trivalent cations like $Bi^{3+}$ together with monovalent cations such as $Ag^{1+}$ into the B-sites of halide perovskites, resulting in the formation of B-cation double perovskites, which follow the general formula $A_2B'B''X_6$. Among these, $Cs_2AgBiBr_6$ and $Cs_2AgBiCl_6$ have emerged as compounds with promising photovoltaic properties and notable stability [11–13]. The experimentally determined optical band gaps for $Cs_2AgBiCl_6$ and $Cs_2AgBiBr_6$ range from 2.20 eV to 2.77 eV and 1.83 eV to 2.19 eV, respectively [12–14]. In comparison, calculations using density functional theory (DFT) and hybrid functionals yield slightly higher band gap values, between 2.62 eV and 3.00 eV for $Cs_2AgBiCl_6$ and 2.06 eV to 2.30 eV for $Cs_2AgBiBr_6$ [12,14]. Although these materials have band gaps in the visible range, they are indirect, making them less suitable for thin-film photovoltaic applications. Zhang et al. [15] proposed that integrating $In^{1+}$ or $Tl^{1+}$ with $Bi^{3+}$ could result in direct band gaps, and Zhao et al. [16] recently reported a direct band gap of 0.91 eV for $Cs_2InBiCl_6$. However, compounds like $Cs_2InBiX_6$ that contain $In^{1+}$ encounter stability issues due to the oxidation of $In^{1+}$ to $In^{3+}$ [17]. S. Mahmud et al. recently studied the lead-free DHP halide of the $A_2AuScX_6$ compound theoretically and found a band gap of 1.30 to 1.93 eV for optoelectronic device and solar cell applications [18]. Transition metals have also been introduced into double halide perovskites (DHPs) due to their unique electronic properties, leading to the synthesis and study of various compounds such as $Cs_2AgCrX_6$ [19,20], $Cs_2AgFeCl_6$ [21], and $Cs_2NaVCl_6$ [22]. Single crystals of $Cs_2Ag_xNa_{1-x}FeCl_6$ and $Cs_2NaSc_{1-x}Cl_6:xTb^{3+}$, have shown promise in photovoltaic and volumetric display technologies [23–25]. Additionally, transition metals from group VIII, including Co, Rh, and Ir, which are commonly found in oxide perovskites, are being studied for their potential in DHPs due to their distinctive optoelectronic properties, which are beneficial for applications in photovoltaic [26–29]. Recent research has renewed interest in Co, Rh, and Ir-based DHPs, with investigations of compounds like $Rb_2NaCoF_6$ [30], $Cs_2AgRhX_6$ [31,32], and $Cs_2CuIrF_6$ [33]. Moreover, studies indicate that the optoelectronic properties of $In^{1+}$-based materials can be improved and stabilized against oxidation by incorporating suitable A-site cations, such as MA and FA organic groups [34]. Certain perovskites are highly stable in air, yet their photovoltaic performance is hindered by their wide band gap values, which are greater than 2.2 eV [35,14]. Recently, V. Deswal *et al.* presented a direct band gap of 1.57 eV DHP $Cs_2AgInBr_6$, achieving a power conversion efficiency (PCE) of 26.9% [36]. Their research on the PCE of the Ag-based DHP, $Cs_2AgInBr_6$, has motivated the search for other Ag-based double perovskites with band gaps close to the Shockley–Queisser (S.Q.) limit, where significant PCE improvements are anticipated. The spectroscopic limited maximum efficiency (SLME) calculates the theoretical maximum efficiency by considering the full solar spectrum, including non-radiative losses, using the principle of detailed balancing [37]. Based on these findings, the combination of $Ag^{1+}$ with $Ir^{3+}$ could potentially meet the requirements for a direct band gap, making it worthy of further investigation for use as perovskite solar cell absorbers.

Acknowledging the paucity of experimental and theoretical research on DHP $A_2AgIrCl_6$ (where A = Cs, Rb, K), we conducted a comprehensive study to evaluate its structural, electronic, optical, mechanical, and thermo-physical characteristics in addition to its stability. These lead-free materials show great promise for photovoltaic applications because of their excellent light absorption coefficients and optimal band gaps. For our analysis, we employed the full-potential linearized augmented plane-wave (FP-LAPW) approach within the DFT framework using Wien2k software. The combination of Ag-Ir cations at the B site with the A and Cl sites provides two key benefits: it allows for the tuning of the band gap from broad to narrow and reduces toxicity compared to lead-based compounds. The efficiency of energy conversion in these materials is evaluated by calculating the spectroscopic limited maximum efficiency. The goal of this study is to provide critical benchmark data that will support and enhance future experimental and theoretical investigations of these compounds. The computational models and techniques used are described in Section 2, and the outcomes and their consequences are covered in Section 3. Section 4 provides an overview of our study's results.

## 2. Computational methods

We investigated the structural, dynamical, mechanical, thermophysical, electronic, and optical properties of the double perovskite $A_2AgIrCl_6$ (A = Cs, Rb, K) in this study by means of first-principles computations. Based on the density functional theory (DFT) framework [38,39], these calculations were carried out using the Wien2k software package's implementation of the full-potential linearized augmented plane-wave (FP-LAPW) approach [40,41]. For the double perovskite bulk structure, we used the exchange-correlation potential due to Perdew-Burke-Ernzerhof (PBE) through the generalized gradient approximation (GGA)[42]. To determine the structural parameters, we applied Birch-Murnaghan's equation of state[43], as illustrated in Fig. 1. The plane wave cut-off parameter $R_{MT} \times k_{max}$ was chosen as 8.0, where $R_{MT}$ denotes the radius of the smallest atomic muffin-tin sphere, and $K_{max}$ represents the maximum wave vector in the plane wave basis. Inside the muffin tin sphere, we specified $l_{max} = 10$ for the maximum partial wave component. The atomic muffin-tin sphere radii values were individually set as $R_{MT}(Cs) = 2.50$ a.u., $R_{MT}(Ag) = 2.13$ a.u., $R_{MT}(Ir) = 2.18$ a.u., and $R_{MT}(X) = 1.88$ a.u. To accurately describe the potential and charge density, we employed a Fourier series with wave vectors up to $G_{max} = 14$ $(Ry^{1/2})$. The computational accuracy was ensured by using a dense $k$-point mesh of 1000 points in the first Brillouin zone (BZ). A cut-off energy of approximately 7.0 Ry separated the valence and core states, while the self-consistent calculation was set to $10^{-5}$ Ry for precise and fully converged energy results. The crystal structure was visualized using VESTA [44]. Moreover, the phonon dynamic stability of the materials was confirmed using the CASTEP (Cambridge Serial Total Energy Package) code [45]. Phonon calculations were conducted on a 1×1×1 supercell model of the unit cell structure. To evaluate lattice dynamic properties such as phonon dispersion and thermodynamic characteristics, we employed the finite displacement supercell approach.

## 3. Results and discussion

### 3.1 Structural parameters and stability

The halide double perovskite materials $A_2AgIrCl_6$ (A = Cs, Rb, K) are depicted in Fig. 1, following a face-centered cubic structure with the $Fm\bar{3}m$ (no. 225) space group [46]. The structure comprises fourteen [IrCl$_6$] octahedrons, thirteen [AgCl$_6$] octahedrons, and eight A (A = Cs, Rb, K) atoms located in the interstitial sites of the octahedrons, contributing to the stabilization of the crystal structure [47]. In the crystal structure of DP $A_2AgIrCl_6$ (A = Cs, Rb, K) within this space group, the unit cell contains $A^{+1}$ cations positioned in the 8$c$ Wyckoff site with the fractional coordinates (0.25, 0.25, 0.25), the $Ag^{+1}$ cations occupy the 4$a$ Wyckoff site with the fractional coordinates (0.5, 0.5, 0.5), $Ir^{+3}$ cations reside in the 4$b$ Wyckoff site precisely at the positions (0.0, 0.0, 0.0) and the $Cl^{-1}$ anion is found at the 24$e$ Wyckoff site precisely at coordinates (0.25, 0.0, 0.0). The HDP $A_2AgIrCl_6$ (A = Cs, Rb, K) compounds have undergone optimization using volume optimization, as illustrated in Fig. 2. The results of this geometry optimization, including the unit cell parameter $a_0$ (Å), bulk modulus $B_0$ (GPa), pressure derivative of $B_0$, and ground state energy $E_0$ (Ry), are summarized in Table 1. The replacement of A-site cation significantly influences the lattice constant of the conventional cell, with the order of Cs, Rb, and K resulting in a decreasing lattice constant. No experimental data are available in the scientific literature to be compared with our findings.

To comprehensively assess the stability of the HDP compounds $A_2AgIrCl_6$ (A = Cs, Rb, K), it is necessary to perform calculations for both their dynamic stability, represented by phonon dispersion relations and their thermodynamic stability. As these calculations demand substantial computational resources and are time-intensive, we initially evaluate the stability of the compounds by computing parameters such as final energy, decomposition enthalpy, formation energy, binding energy, Goldschmidt's tolerance factor, octahedral factor, and a new tolerance factor. The final energy values per atom for $Cs_2AgIrCl_6$, $Rb_2AgIrCl_6$, and $K_2AgIrCl_6$ were determined to be -2149.64 eV, -2004.61 eV, and -2023.32 eV, respectively.

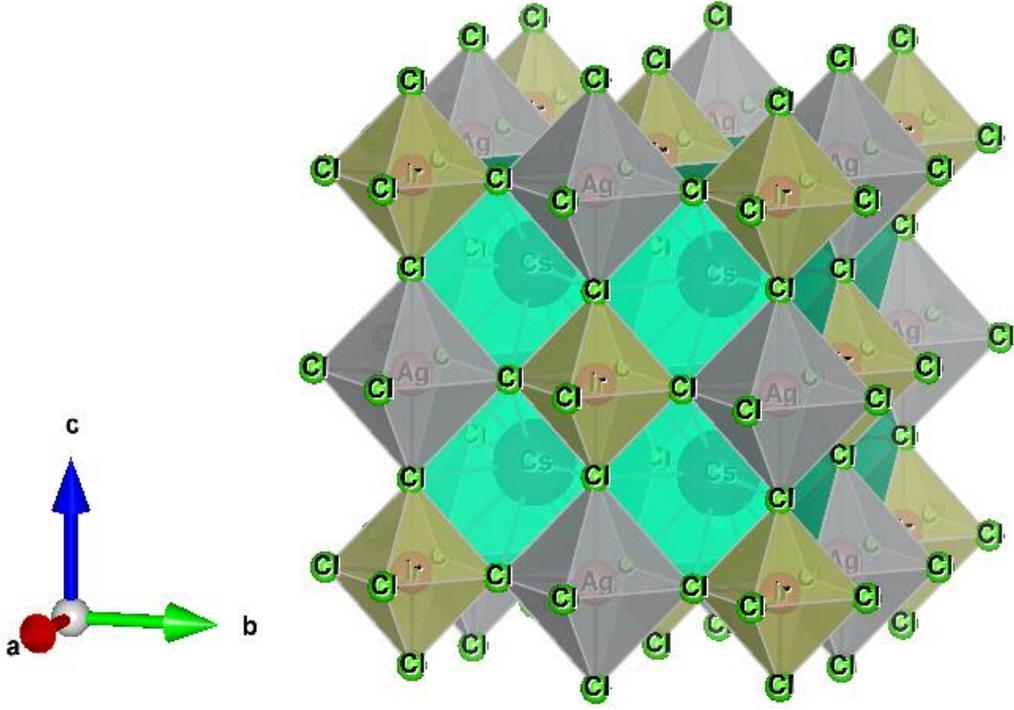

Fig. 1. The unit cell of $Cs_2AgIrCl_6$.

To evaluate the thermodynamic stability of double halide $A_2AgIrCl_6$ (A = Cs, Rb, K) perovskites, we compute their decomposition energy along different potential pathways. The primary and most significant pathway involves the decomposition of $A_2AgIrCl_6$ (A = Cs, Rb, K) into the respective binary materials. Typically, halide perovskites are synthesized through the inverse reactions of these binary materials. Specifically, we determine the decomposition energy, which is defined as follows:

$$\Delta H_D = 2E[ACl] + E[AgCl] + E[IrCl_3] - E[A_2AgIrCl_6] \qquad (1)$$

Where, $E[ACl]$, $E[AgCl]$, $E[IrCl_3]$ and $E[A_2AgIrCl_6]$ are the final energies of $ACl$, $AgCl$, $IrCl_3$ and $A_2AgIrCl_6$, respectively [48]. The computed $\Delta H_D$ values are given in Table 1. The significant positive $\Delta H_D$ indicates that energy is gained from the decomposed products, confirming the thermodynamic stability of all the studied materials. Moreover, for examining the thermodynamic stability of perovskite materials, one can determine the formation energy ($E_f$) and binding energies ($E_B$) through the formula mentioned [49]:

$$E_f = \frac{E_{A_2AgIrCl_6} - n_A \times E_A - n_{Ag} \times E_{Ag} - n_{Ir} \times E_{Ir} - n_{Cl} \times E_{Cl}}{40} \qquad (2)$$

$$E_B = \frac{E_{A_2AgIrCl_6} - n_A \times \mu_A - n_{Ag} \times \mu_{Ag} - n_{Ir} \times \mu_{Ir} - n_{Cl} \times \mu_{Cl}}{40} \qquad (3)$$

In this context, $E_{A_2AgIrCl_6}$ signifies the overall energy of double halide perovskite materials, while $E_A$, $E_{Ag}$, $E_{Ir}$ and $E_{Cl}$ represent the energies of individual A (A = Cs, Rb, K), Ag, Ir, and Cl atoms, respectively. The variable '$n$' denotes the number of atoms and '$\mu$' symbolizes the free energy of each atom. The computed formation energies ($E_f$) and binding energies ($E_B$) for these three perovskite materials are presented in Table 1. These values exhibit negativity, indicating that these three perovskites should be synthesizable and they comply with thermodynamic stability criteria.

In most cases, the crystallographic stability of the perovskite structure can be anticipated using the Goldschmidt tolerance factor denoted as '$t$' in equation (4) [50], as well as the octahedral factor '$\eta$' presented in equation (5) [51]. Recently, Bartel and co-workers introduced a new tolerance factor denoted as '$\tau$' in equation (6) [52], which has demonstrated a high level of predictive accuracy. The parameters $t$, $u$, and $\tau$ are determined using the following equations:

$$t = \frac{(R_A + R_X)}{\sqrt{2}(R_B + R_X)} \tag{4}$$

$$u = \frac{R_B}{R_X} \tag{5}$$

$$\tau = \frac{R_X}{R_B} - n_A\left(n_A - \frac{R_A/R_B}{\ln(R_A/R_B)}\right) \tag{6}$$

In these equations, $n_A$ denotes the oxidation state of A, by definition, $R_A > R_B$, and $R_A$, $R_B$, and $R_X$ correspond to the ionic radii of A, B, and X ions, respectively in an ABX$_3$ perovskite structure. Due to the presence of a double perovskite structure with distinct radii for B$^+$ and B$^{3+}$ sites, we calculate the mean of the ionic radii of Ag$^+$ and Ir$^{3+}$ to determine $R_B$. To establish a perovskite structure with stability, it is necessary that the Goldschmidt tolerance factor $t$ resides within the interval of 0.81 to 1.11, the octahedral factor $\mu$ falls within the range of 0.41 to 0.90, and the new tolerance factor $\tau$ must not exceed 4.18 [50–52]. We assessed these parameters by employing Shannon's ionic radii [53]. The Goldschmidt tolerance factor, octahedral factor, and new tolerance factor have been computed and are presented in Table 2.

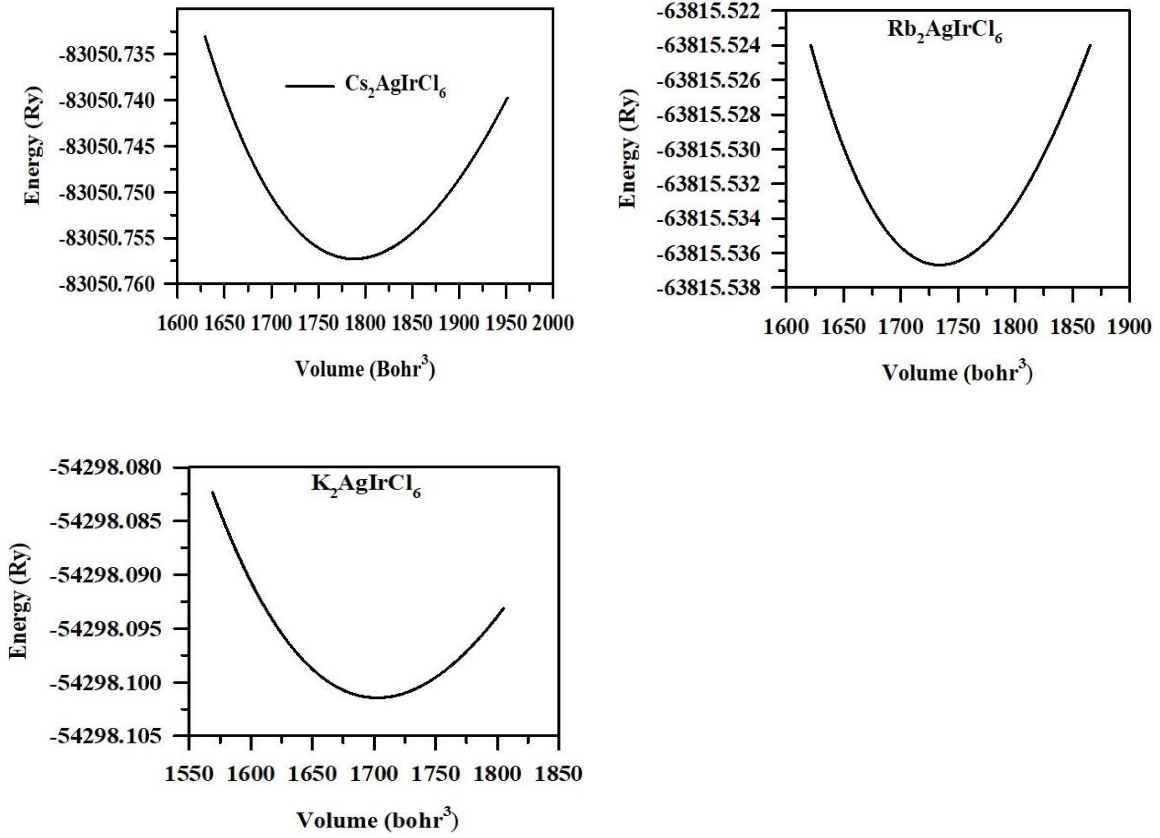

Fig. 2. Birch–Murnaghan total energy vs. volume fitting curve for the $A_2AgIrCl_6$ (A = Cs, Rb, K) compounds.

**Table 1:** The lattice constant $a_0$ (Å), bulk modulus $B_0$ (GPa), first derivative of bulk modulus $B_0^{'}$, total energy $E_{tot}$ (Ry), decomposition energy $\Delta H_D$ (meV/atom), formation energy $E_f$ (eV/atom) and binding energy $E_B$ (eV/atom) for double perovskite (DP) $A_2AgIrCl_6$ (A = Cs, Rb, K).

| DP | $a_0$ (Å) | $B_0$ (GPa) | $B_0^{'}$ | $E_{tot}$ (Ry) | $\Delta H_D$ (meV/atom) | $E_f$ (eV/atom) | $E_B$ (eV/atom) |
|---|---|---|---|---|---|---|---|
| **$Cs_2AgIrCl_6$** | 10.19 | 41.53 | 5.32 | -83050.76 | 73.55 | -2.10 | -4.29 |
| **$Rb_2AgIrCl_6$** | 10.09 | 44.03 | 5.60 | -63815.54 | 50.66 | -2.06 | -4.28 |
| **$K_2AgIrCl_6$** | 10.03 | 44.93 | 5.47 | -54298.10 | 0.19 | -2.03 | -4.28 |

**Table 2:** Shannon's ionic radii ($r$) of ions, Gold-Schmidt tolerance factor ($t$), octahedral factor ($\mu$), and new tolerance factor ($\tau$) for $A_2AgIrCl_6$ (A = Cs, Rb, K) compounds.

| DP | Ionic radius of cations (Å) | | Ionic radius of anion (Å) | Tolerance factor ($t$) | Octahedral factor ($u$) | New tolerance factor ($\tau$) |
|---|---|---|---|---|---|---|
| $Cs_2AgIrCl_6$ | $r_{Cs}$ 1.88 | $(r_{Ag} + r_{Ir})/2$ 0.92 | $r_{Cl}$ 1.81 | 0.96 | 0.51 | 3.83 |
| $Rb_2AgIrCl_6$ | $r_{Rb}$ 1.72 | $(r_{Ag} + r_{Ir})/2$ 0.92 | $r_{Cl}$ 1.81 | 0.91 | 0.51 | 3.93 |
| $K_2AgIrCl_6$ | $r_K$ 1.64 | $(r_{Ag} + r_{Ir})/2$ 0.92 | $r_{Cl}$ 1.81 | 0.89 | 0.51 | 4.04 |

### 3.2 Phonon stability

Figure 3 displays the phonon dispersion curves (PDCs) and phonon density of states (PDOS) for $A_2AgIrCl_6$, where A represents Cs, Rb, or K. A dynamics matrix, derived from force constants, illustrates the variation in force experienced by a standard atom due to the orientation of neighboring atoms. The matrices' diagonals are then isolated, representing eigen-values and eigenvectors, indicative of fundamental phonon frequencies and motion. The results reveal three acoustic and multiple optical phonon modes, consistent with the dispersion curves. However, upon deploying conventional unit cells, it was found that two modes were degenerate, underscoring the importance of cell structure in determining degeneracy accurately [32].

The conventional unit cell of $A_2AgIrCl_6$ (where A = Cs, Rb, K) comprises forty atoms, resulting in a variety of acoustic and optical phonon modes. Among these, three acoustic modes at the Γ point are characterized by low frequencies ranging from 0 to 1.25 THz, while the remaining optical modes fall within the high-frequency range of 1.25 to 10 THz. Figure 3 indicates the absence of soft phonon modes in the material, as all phonon modes exhibit positive lattice vibration frequencies. This confirms the dynamic stability of the material. The phonon stability of $A_2AgIrCl_6$ (A = Cs, Rb, K) is maintained along the pathways W to L, L to Γ, Γ to X, X to W, and W to K, ensuring stability across the entire pathway. The PDOS of the compound is presented alongside the dispersion curve to enhance the clarity of analysis. Both the PDOS and the phonon dispersion have been meticulously adjusted to capture these effects. It is shown that lattice vibrations at low frequencies, less than 1.0 THz, correspond to optical modes and are primarily impacted by A-site atoms, namely Cs, Rb, and K. On the other hand, the low and mid-frequency vibrations are mostly controlled by Ag, Ir, and Cl atoms. Higher frequencies primarily affect the vibrational behavior due to the interaction between Ir and Cl ions. It can be inferred from the phonon density of state profiles that lattice thermal conductivity and the low-T heat capacity of these compounds are dominated by the vibrational modes originating from the Cl atoms.

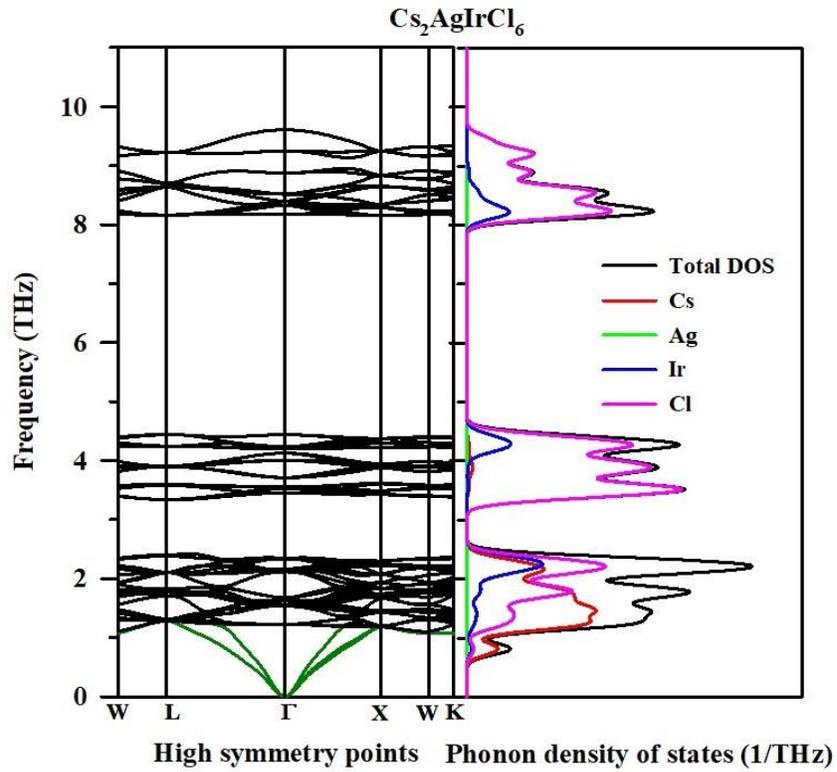
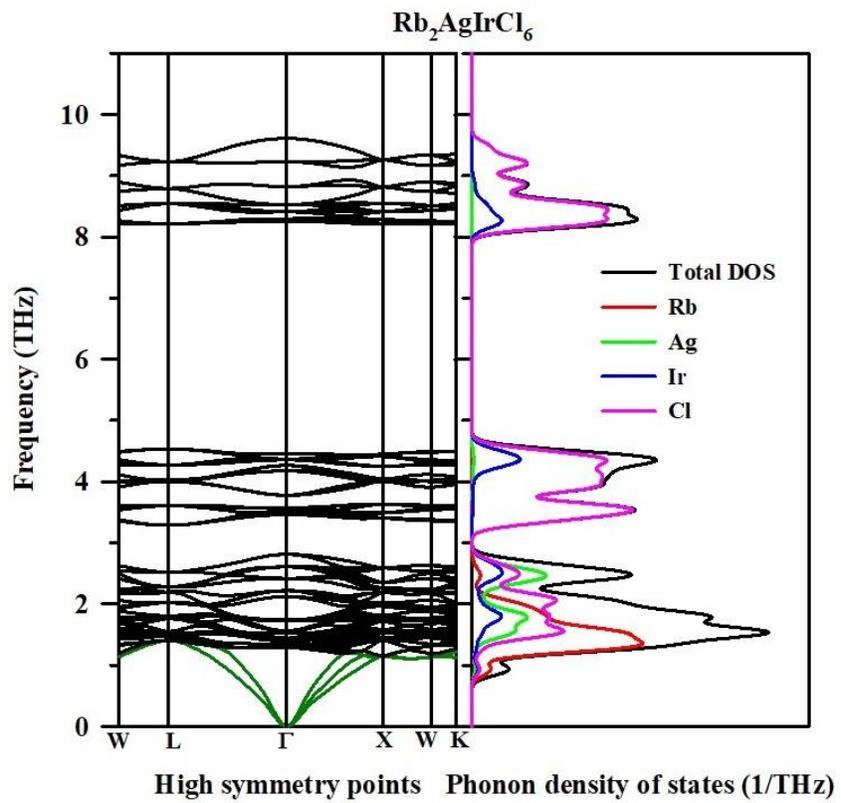

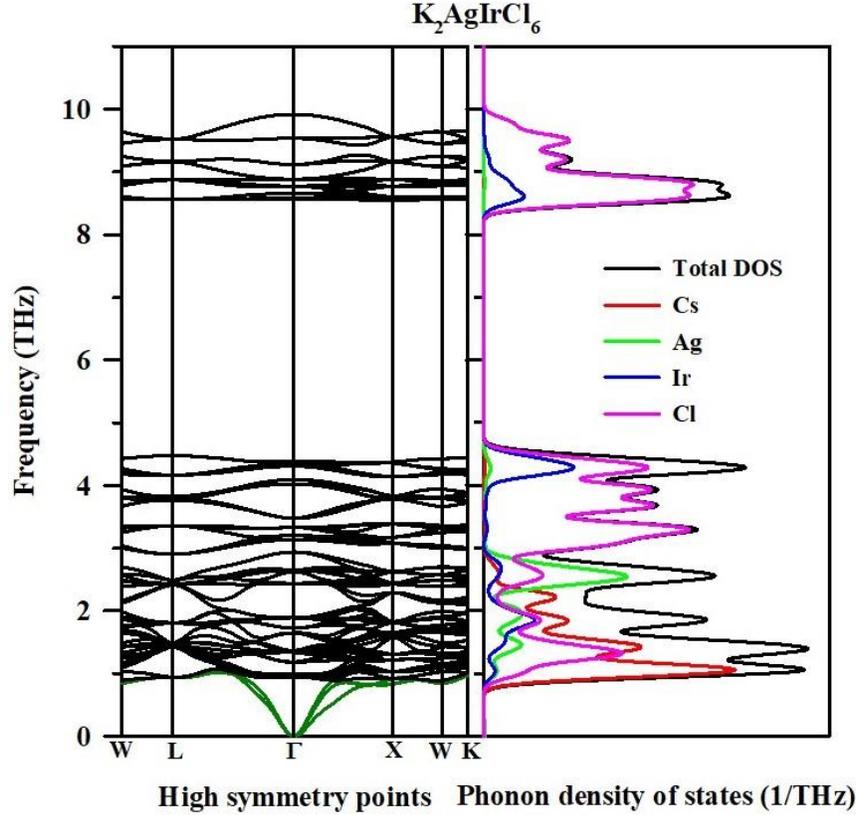

Fig. 3. Phonon dispersion curve along with phonon DOS of $A_2AgIrCl_6$ (A = Cs, Rb, K) compounds.

## 3.3 Electronic properties

The features of direct bandgap semiconductors make them more favorable for optical applications than those with indirect bandgaps [54]. The GGA-calculated bandgap ($E_g$) for $Cs_2AgIrCl_6$ is 0.34 eV, suggesting the potential for electronic transitions from the valence band maximum (VBM) to the conduction band minimum (CBM). Substituting lighter alkali atoms (A = Cs, Rb, K) at the A-site in $A_2AgIrCl_6$ has a minimal impact on $E_g$, with values ranging from 0.34 to 0.38 (see Table 3). In contrast, the TB-mBJ functionals yield much higher $E_g$ values of 1.43 eV, 1.50 eV, and 1.55 eV for $Cs_2AgIrCl_6$, $Rb_2AgIrCl_6$, and $K_2AgIrCl_6$ (see Table 3), respectively. This mismatch shows that compared to the TB-mBJ functional, GGA functionals tend to significantly underestimate $E_g$ while maintaining the direct nature of the bandgap transition between VBM and CBM. The GGA functional underestimates the band gap [55–57], while the TB-mBJ functional provides results that closely match the experimental values documented in the literature [58–60]. Both approximations consistently yield a direct band nature at the same symmetry points. Fig. 4 illustrates band structure computations for $A_2AgIrCl_6$ (A = Cs, Rb, K) using GGA (left) and TB-mBJ (right), revealing a significant behavioral similarity. Our discussion exclusively focuses on TB-mBJ for its superior accuracy in energy band gap values over GGA. Perovskite materials with band gap values between 0.8 and 2.2 eV [61] are highly suitable for a wide range of photoelectric applications, particularly in photoelectric conversion processes. The double perovskites $A_2AgIrCl_6$ (A = Cs, Rb, K), with their band gaps falling within this ideal range, carry significant potential for development as photosensitive materials for future solar cell technologies. Their band gap properties make them promising candidates for enhancing the efficiency and effectiveness of photoelectric conversion in solar energy applications.

Analyzing the density of states (DOS) and electronic band structure plots for the $A_2AgIrCl_6$ (A = Cs, Rb, K) compounds (see Fig. 4 and 5), our investigation reveals that the valence band maximum (VBM) predominantly comprises the character of Cl(3P) and Ir(5d). Specifically, the dispersion of the valence band just below the Fermi level is intricately influenced by the non-bonding orbital states $t_{2g}$ of Ir and the 3p orbital states of Cl. In the case of $Cs_2AgIrCl_6$, the calculated normalized contribution to the VBM is 65.2% for Ir(5d) and 20.6% for Cl(3p). Similarly, for $Rb_2AgIrCl_6$ (and $K_2AgIrCl_6$), the contributions to the VBM for Ir(5d) and Cl(3p) are 65.0 (64.5)% and 19.2 (17.9)%, respectively. Interestingly, the alkali and Ag atoms demonstrate minimal contributions, hovering around 1-2%, to the VBM for $A_2AgIrCl_6$ (A = Cs, Rb, K). Conversely, our examination of the conduction band minimum (CBM) reveals a predominant derivation from Ir(5d) empty anti-bonding states $e_g$, with significant contributions from Ag, Ir, and Cl states, leading to its dispersion well above the Fermi level. The normalized contributions to the CBM for $A_2AgIrCl_6$ (A = Cs, Rb, K) are (57.2, 57.1, 56.8)% for Ir(5d), (27.1, 27.0, 26.8)% for Cl(3p), and (5.8, 5.8, 6.0)% for Ag(4d), respectively. Notably, these contributions exhibit marginal variations upon the substitution of the A-site cation with lighter alkali group elements. Considering the substantial separation between the CBM and VBM, coupled with the VBM's proximity to the Fermi level, our analysis leads to the inference that $A_2AgIrCl_6$ (A = Cs, Rb, K) can be characterized as p-type materials [62]. Very large DOS peaks just below the Fermi level come from the nearly localized Ir(5d) electrons.

Studying the effective mass of holes and electrons is essential for gaining a deeper understanding of the photovoltaic properties of solar cells, as it significantly impacts carrier mobility, resistivity, and the optical response of free carriers [63]. We observed that the top of the valence band is less dispersed compared to the bottom of the conduction band. This similarity in the flatness of the valence band was also noted in bulk $Cs_2AgInCl_6$, originating from Ag 4d and Cl 3p orbitals [73]. In such cases, it's common for the hole effective mass ($m_h^*$) linked with the valence band maximum (VBM) to be greater than the electron effective mass ($m_e^*$) of the conduction band maximum (CBM). Consequently, the carrier mobility of holes is typically lesser than that of electrons. This curvature is inversely related to the second derivative of energy concerning the wave vector $k$ and is elucidated by the dispersion relationship [64]:

$$m^* = \pm \frac{\hbar^2}{(d^2 E(k)/dk^2)} \quad (7)$$

In this context, the + and – signs indicate electrons and holes, respectively. The symbols $m^*$, $k$, $E(k)$, and $\hbar$ denote the effective mass of the electron or hole, the wave vector, the energy as a function of $k$, and the reduced Planck's constant, respectively. The values of $d^2E(k)/dk^2$ are obtained from *E-k* dispersion curve through fitting a parabolic curve at symmetry points. Table 3 provides the estimated effective masses of electrons and holes of the studied compounds. For $Cs_2AgIrCl_6$, the electron's effective mass is approximately 0.19 $m_e$ and the hole's is 1.43 $m_e$, where $m_e$ is the mass of a free electron. For $Rb_2AgIrCl_6$, these values are 0.18 $m_e$ and 1.70 $m_e$, respectively; and for $K_2AgIrCl_6$, they are 0.18 $m_e$ and 1.14 $m_e$. All these values were computed using the TB-mBJ functionals. The electron effective masses are all less than $m_e$, while the hole effective masses exceed $m_e$. Table 3 reveals that the hole effective masses ($m_h^*$) are significantly greater than the electron effective masses ($m_e^*$), indicating that these materials function as hole-transporting (p-type) semiconductors [65]. Additionally, Table 3 shows that both carriers have low effective masses, suggesting higher carrier mobility ($\mu$) since $m^*$ and $\mu$ are inversely related [66]. This suggests that the double perovskites $A_2AgIrCl_6$ (A = Cs, Rb, K) are highly promising as photovoltaic materials, with significant potential for a wide range of photovoltaic applications.

**Table 3:** Band gaps, band gap nature, and calculated electron and hole effective mass values of DP $A_2AgIrCl_6$ (A = Cs, Rb, K) compounds.

| Compounds | Band gap (eV) | Functional | Band gap nature | Effective mass of electron, $m_e^*$ | Effective mass of hole, $m_h^*$ |
|---|---|---|---|---|---|
| $Cs_2AgIrCl_6$ | 0.34 | GGA PBE | Direct | 0.13 $m_e$ | 1.10 $m_e$ |
|  | 1.43 | Tb-mBJ | Direct | 0.19 $m_e$ | 1.43 $m_e$ |
| $Rb_2AgIrCl_6$ | 0.36 | GGA PBE | Direct | 0.13 $m_e$ | 1.10 $m_e$ |
|  | 1.50 | Tb-mBJ | Direct | 0.18 $m_e$ | 1.70 $m_e$ |
| $K_2AgIrCl_6$ | 0.38 | GGA PBE | Direct | 0.13 $m_e$ | 1.64 $m_e$ |
|  | 1.55 | Tb-mBJ | Direct | 0.18 $m_e$ | 1.14 $m_e$ |

*$m_e$ = mass of electron

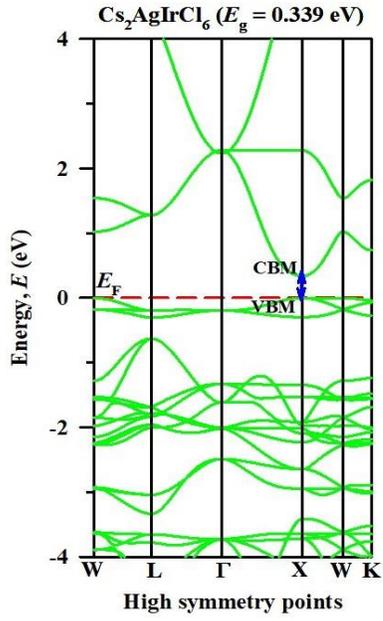 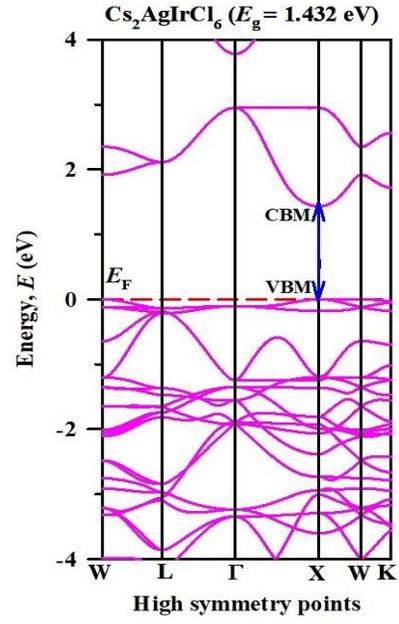
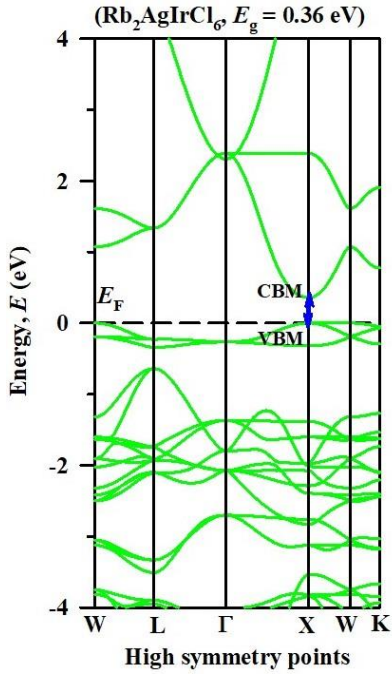 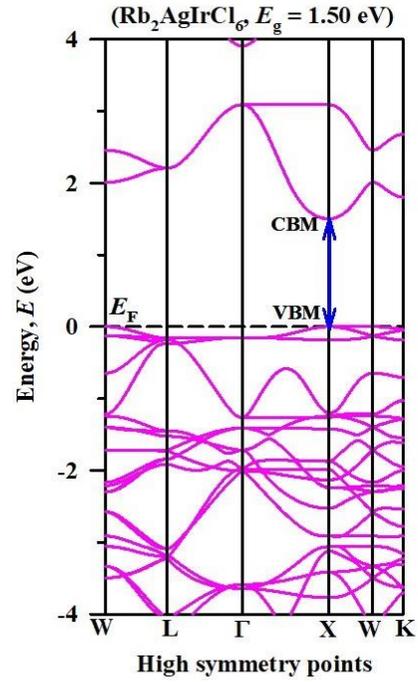

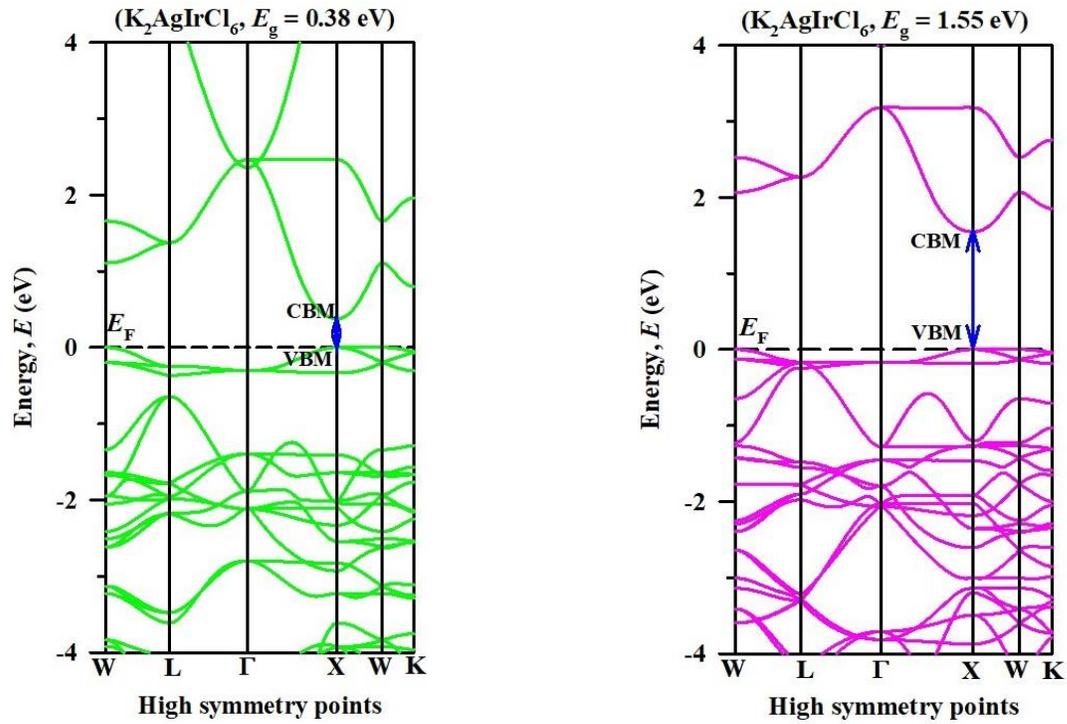

Fig. 4. Band structure of $A_2AgIrCl_6$ (A = Cs, Rb, K) compounds using GGA (left panel, green color) and TB-mBj (right panel, pink color) functionals.

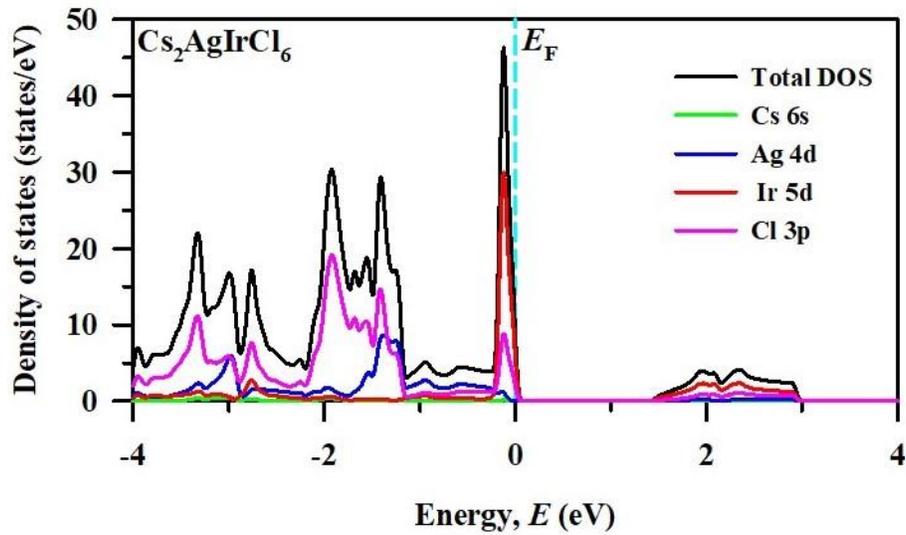

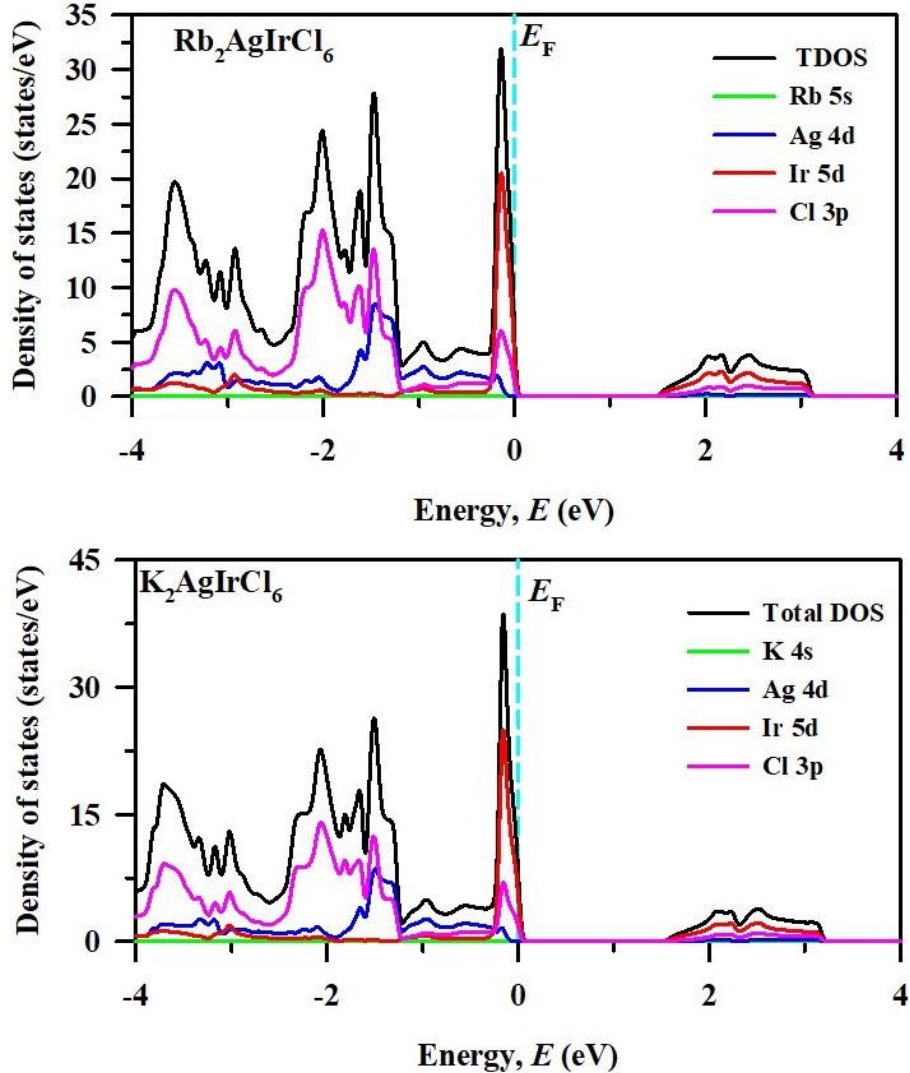

Fig. 5. Density of states of $A_2AgIrCl_6$ (A = Cs, Rb, K) compounds.

### 3.4 Charge density

$A_2AgIrCl_6$ compounds (A = Cs, Rb, K) are part of the inorganic DP family, showcasing unique bonding patterns and structures influenced by $Cs^+/Rb^+/K^+$ cations. Analyzing charge density and bonding in relation to crystal structure and electronic interactions, we found that these compounds exhibit a cubic crystal structure (Fm-3m space group). The arrangement involves Cl- ions at central octahedral positions, with Ag and Ir ions at corners, sharing octahedral positions. Interstitial spaces between $Cs^+/Rb^+/K^+$ ions form a network of corner-sharing octahedra, impacting electronic and bonding characteristics. The A ions (Cs/Rb/K) engage in ionic bonding with Cl- ions due to their higher electronegativity, forming A-Cl bonds. Covalent bonding in the Ag-Ir-Cl octahedral and potential metallic character in Ag-Ir bonding contribute to the materials' charge density distribution. In summary, the intricate interplay of ionic, covalent, and metallic bonding within the crystal structure defines the charge density and bonding nature of $A_2AgIrCl_6$.

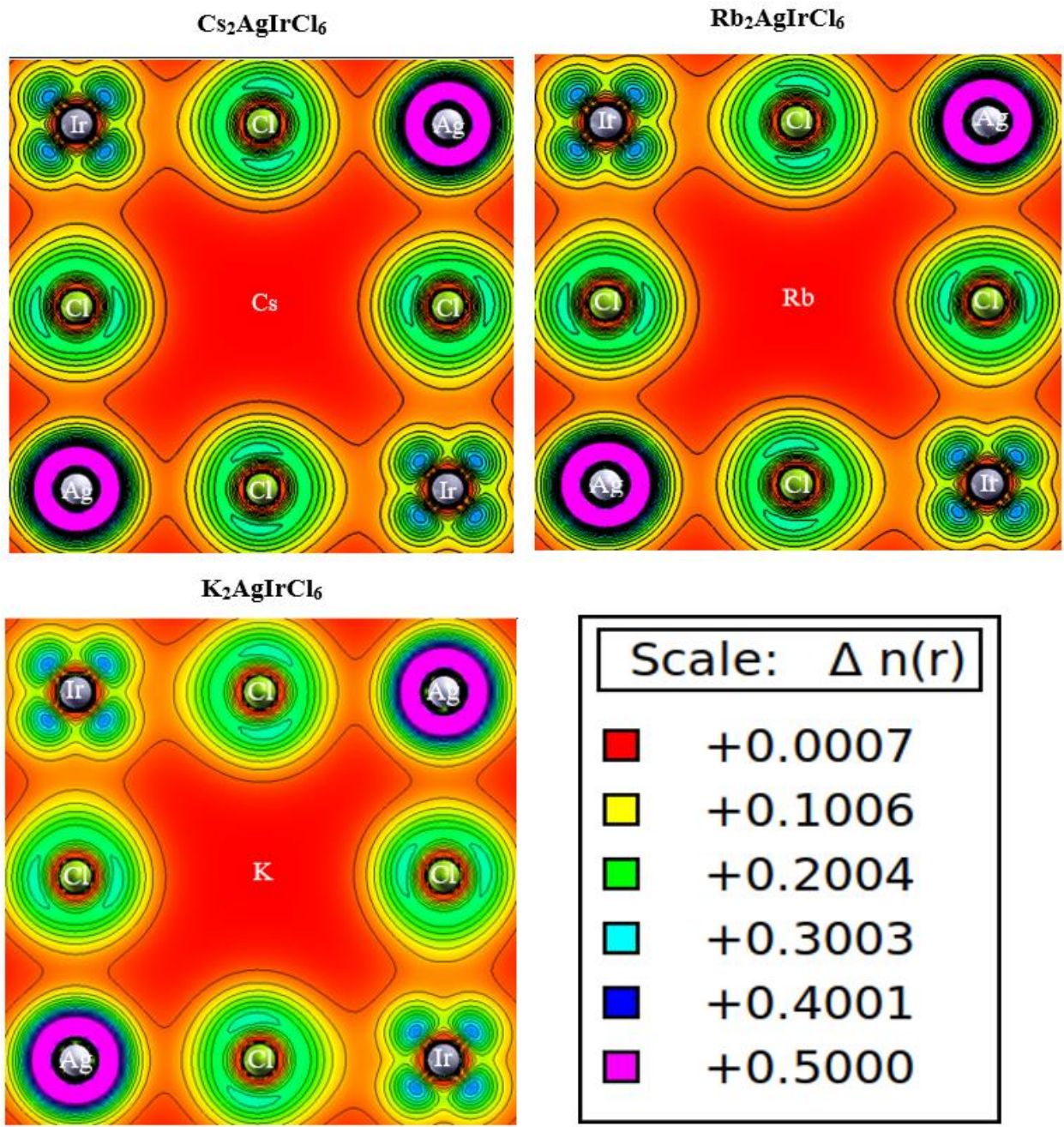

Fig. 6. Charge density of $A_2AgIrCl_6$ (A = Cs, Rb, K) compounds.

## 3.5 Optical properties

Halide perovskites are recognized for their distinctive optical characteristics, such as robust absorption in the visible and UV regions of the solar spectrum, along with a high optical absorption coefficient, distinguishing them from conventional solar absorbers. In the previous section, we have found the direct band gap values of 1.43 eV, 1.50 eV, and 1.55 eV for $Cs_2AgIrCl_6$, $Rb_2AgIrCl_6$, and $K_2AgIrCl_6$, respectively which clearly demonstrated a promising prospect for solar cell applications. The optical attributes, specifically concerning light absorption and emission, hold significant importance for optoelectronic solid materials. Consequently, our primary focus is on evaluating the optical properties of $A_2AgIrCl_6$ (A = Cs, Rb, K). The fundamental real and imaginary components of the dielectric function, $\varepsilon_1(\omega)$ and $\varepsilon_2(\omega)$, delineate the frequency-dependent electronic transition features of the band structure. The complex dielectric function encapsulates the complete material response to electromagnetic radiation-induced perturbations [67], expressed by the equation $\varepsilon(\omega) = \varepsilon_1(\omega) + i\varepsilon_2(\omega)$ [68]. In Fig. 7(a) and (b), we compared the spectra of $\varepsilon_1(\omega)$ and $\varepsilon_2(\omega)$ for $A_2AgIrCl_6$ perovskites within the energy range of 0 eV to 14 eV. The static dielectric function $\varepsilon_1(0)$ assumes values of 4.41, 4.27, and 4.20 for $Cs_2AgIrCl_6$, $Rb_2AgIrCl_6$, and $K_2AgIrCl_6$, respectively, mirroring those of $Cs_2AgRhCl_6$ (A = Li, K, Na, Rb, Cs) [32], $Cs_2XRhCl_6$ (X = Na, K) [69], and $Cs_2AgBiCl_6$ [70]. Substitution of heavier alkali metals results in a higher value. Additionally, $\varepsilon_1(\omega)$ rises with ascending photon energy for each perovskite material, reaching maximum values of 10.84, 10.76, and 10.55 at energies 2.24 eV, 2.33 eV, and 2.38 eV for $Cs_2AgIrCl_6$, $Rb_2AgIrCl_6$, and $K_2AgIrCl_6$, respectively. Following this, there is a rapid decline in $\varepsilon_1(\omega)$ characterized by oscillations and multiple peaks. In certain intervals, $\varepsilon_1(\omega)$ records values below zero, indicating an impediment for photons with corresponding energies to penetrate the solid materials. Penn's model establishes an inverse relationship between static polarization and band gap given by[71]:

$$\varepsilon_1(0) \approx 1 + \left[\frac{\hbar\omega_p}{E_g}\right]^2 \quad (8)$$

Here, $\hbar$ and $\omega_p$ represent the reduced Planck constant and the plasma frequency, respectively.

The significance of $\varepsilon_2(\omega)$ lies in its pivotal role in determining the maximum absorption and overseeing interband transitions within materials. Due to inherent limitations in DFT, there are slight deviations in the transition points (VB to CB) from the band structure, as illustrated in Fig.7(b). The peak value of $\varepsilon_2(\omega)$ is commonly denoted as the first absorption peak (FAP) and is linked to electronic transitions at the Fermi level. For instance, in the case of $A_2AgIrCl_6$ (A= Cs/Rb/K), the FAP is observed at 2.38, 2.46, and 2.52 eV. These are situated at energies higher than those observed in the $\varepsilon_1(\omega)$ curves (refer to the above). The primary focus for these compounds is on the highest FAP values. These values fall within the visible range, indicating an effective absorption capacity for visible light. In comparison to the DOS, the first absorption spectrum (FAS) primarily stems from the Ir-d and Cl-p orbitals at the valence band maximum (VBM) and the conduction band minimum (CBM) points.

The refractive factor n(ω), extinction factor k(ω), absorption coefficient α(ω), conductivity σ(ω), reflectivity R(ω), and loss factor L(ω), which are essential optical constants, are derived from the dielectric properties of all compounds using prescribed equations given below [72,73]:

$$n(\omega) = \left[\frac{\varepsilon_1(\omega)}{2} + \frac{\sqrt{\varepsilon_1^2(\omega) + \varepsilon_2^2(\omega)}}{2}\right]^{1/2} \qquad (9)$$

$$k(\omega) = \left[\frac{-\varepsilon_1(\omega)}{2} + \frac{\sqrt{\varepsilon_1^2(\omega) + \varepsilon_2^2(\omega)}}{2}\right]^{1/2} \qquad (10)$$

$$\alpha(\omega) = \frac{2\omega}{c} k(\omega) \qquad (11)$$

$$\sigma(\omega) = \frac{4\varepsilon_0 E}{e} \varepsilon_2(\omega) \qquad (12)$$

$$R(\omega) = \frac{(n-1)^2 + k^2}{(n+1)^2 + k^2} \qquad (13)$$

$$L(\omega) = \frac{\varepsilon_2(\omega)}{\varepsilon_1^2(\omega) + \varepsilon_2^2(\omega)} \qquad (14)$$

The complex refractive index $(n + ik)$ is a vital property of solid materials, providing insights into the speed of light propagation and potential applications in optoelectronics [74]. Using equations (9) and (10), we determined the real components of the complex refractive index. The value of $n(\omega)$ varies depending on the material, while semiconductors typically have a small $k(\omega)$ [75]. The outcomes of the refractive index calculation are presented in Fig.7(c). The correlation between $n^2(\omega) = \varepsilon_1(\omega)$ resulted in a parallel behavior of the real part of the dielectric function and the refractive index [51]. The static refractive index at zero frequency for $A_2AgIrCl_6$ (A = Cs, Rb, K) is respectively 2.09, 2.06, and 2.04. The studied materials have $n(\omega)$ values within the optoelectronic suitability range of 2.0 to 4.0 [35]. The peak values of $n(\omega)$ at energies of 2.27, 2.35, and 2.41 eV for $A_2AgIrCl_6$ (A = Cs, Rb, K) are 3.44, 3.43, and 3.40. These values are suitable for optical ambient materials with typical refractive indices of 2.5-3.5 [74]. The extinction coefficient $k(\omega)$ exhibits a consistent resemblance to the patterns of $\varepsilon_2(\omega)$ and $\alpha(\omega)$, indicating the detection of electromagnetic wave loss in materials, as illustrated in Fig. 7(d). The relationship between $k(\omega)$ and $\alpha(\omega)$ is as follows: $k = \frac{\alpha\lambda}{4\pi}$.

The efficiency of optoelectronic and photovoltaic devices is largely influenced by their ability to absorb light, as the generation and transport of charge carriers are directly tied to absorption coefficient [76]. The absorption coefficient, $\alpha(\omega)$, in semiconductors plays a crucial role in explaining how materials absorb photons and produce electron-hole pairs, which is fundamental to the functionality of photovoltaic devices like solar cells. A higher absorption coefficient leads to more effective light absorption and a greater reduction in the intensity of light passing through the material. This ultimately enhances the device's performance. Additionally, the absorption coefficient $\alpha(\omega)$ is similar to $\varepsilon_2(\omega)$, as shown by their coefficients at various energies and wavelengths in Fig. 7(e) and 7(f). The absorption edge, also known as the optical band gap, correlates precisely with the electronic band gap. Specifically, for $A_2AgIrCl_6$ (A = Cs, Rb, K), the curve displays initial peaks at 2.46, 2.57, and 2.62 eV, respectively, all within the visible range. This emphasizes the effective use of these double halide perovskites as absorbents for the visible spectrum, suggesting their potential utility in photovoltaic and photodetector applications[77]. The existence of multiple peaks in the higher range (6.0 to 13.5 eV) with varying intensities signifies diverse transitions from filled to unfilled states. The visible range $\alpha(\omega)$ values at the highest peaks measure approximately $4.9 \times 10^5$, $5.1 \times 10^5$, and $5.2 \times 10^5$ cm$^{-1}$ for $A_2AgIrCl_6$ (A = Cs, Rb, K), respectively. This indicates a significant absorption of visible light by these systems. The absorption coefficients of our investigated materials surpass those of others, ranging from 6.12 to $6.58 \times 10^4$ cm$^{-1}$ for $A_2CuSbX_6$ (A = Cs, Rb, K; X = Cl, Br, I) [78], 3 to $6.5 \times 10^4$ cm$^{-1}$ for $Cs_2CuBiX_6$ (X = Cl, Br, I) [79], 2.3 $to$ $3.4 \times 10^5$ cm$^{-1}$ for $Cs_2AgBiX_6$ (X = Cl, Br, I) [80]. They are comparable to those of $Cs_2XRhCl_6$ [69] and $Cs_2AgRhCl_6$ [32] (around $10^5$), as observed in the highest peaks of double halide perovskites compounds within the visible range (400-700 nm). As shown in Fig. 7(f), the absorption peaks for the double perovskites $A_2AgIrCl_6$ (A = Cs, Rb, K) are observed at 498 nm, 482 nm, and 477 nm, respectively. These peaks all fall within the visible light spectrum of 380–780 nm, indicating that these perovskite materials have strong absorption capabilities for visible light. The optical band gap in the studied double halide perovskites is determined through the established Tauc plot from absorption coefficient spectra [81]. Accurate estimation of the optical band gap is crucial for predicting semiconductor properties. Misuse of the Tauc plot may lead to misinformation, especially in estimating the band gap [82]. The Tauc equation, based on the absorption coefficient, is expressed as $(\alpha h\nu)^{1/\eta} = A(h\nu - E_g)$, where, $\alpha$, $h$, $\nu$, $A$, and $E_g$ represent the absorption coefficient, Planck constant, frequency, a material-dependent coefficient, and the band gap, respectively [81]. The parameter η varies based on the bandgap nature, typically using 1/2 and 2 for direct and indirect band gap semiconductors. Fig. 7(f) illustrates the optical band gap estimation for $A_2AgIrCl_6$ (A = Cs, Rb, K) as 1.40, 1.45, and 1.48 eV, respectively, using the TB-mBJ functional, aligning with electronic band structure gaps shown in Fig. 4 and Table 3.

Optical conductivity, described by equation (12), governs electron conduction in materials when exposed to specific photon frequencies. Optoelectronic devices require the optical conductivity, linked to interband electron movement, to be between 1.4 eV and 4.0 eV in the visible spectrum [35]. Alkali metal substitution has minimal impact on conductivity. The primary conductivity peaks for $Cs_2AgIrCl_6$, $Rb_2AgIrCl_6$, and $K_2AgIrCl_6$ occur at 2.38 eV (3534 1/Ω-cm), 2.46 eV (3691 1/Ω-cm), and 2.51 eV (3699 1/Ω-cm), respectively, with the highest peak observed in the range of 5200 to 5500 1/Ω-cm from 11 to 13 eV, as illustrated in Fig. 7(g). These peaks, predominantly influenced by Ir-5d states, play a crucial role in the material's overall electronic structure. These findings affirm the high optical conductivity of these materials in the visible region.

The reflectance spectra of $A_2AgIrCl_6$ double halide perovskite compounds (where A = Cs, Rb, K) are analyzed in Fig. 7(h). Calculated static reflection coefficients $R(0)$ for DP are 0.125, 0.121, and 0.118 for $A_2AgIrCl_6$ (A = Cs, Rb, K). Minimal reflectivity $R(\omega)$ suggests efficient light capture within the energy range from zero to the band gap. $R(0)$ increases with the substitution of heavier alkali atoms, and peak positions around 2.5 eV for $A_2AgIrCl_6$ (A = Cs, Rb, K) correspond to a maximum reflectance of about 38%. Fig. 7(i) displays the energy loss function, representing the electrons' energy loss as they move. Notable peaks at approximately 3.52 eV ($Cs_2AgIrCl_6$), 3.68 eV ($Rb_2AgIrCl_6$), and 3.76 eV ($K_2AgIrCl_6$) indicate electron dispersion around these energies, correlating with optical conductivity. Materials with low reflectivity and loss function values are valuable for potential use in solar cell absorbing layers. These materials, featuring an adaptable band gap, robust dielectric strength, impressive absorbance spectra, and favorable photoconductivity, demonstrate promise for applications in optoelectronic devices.

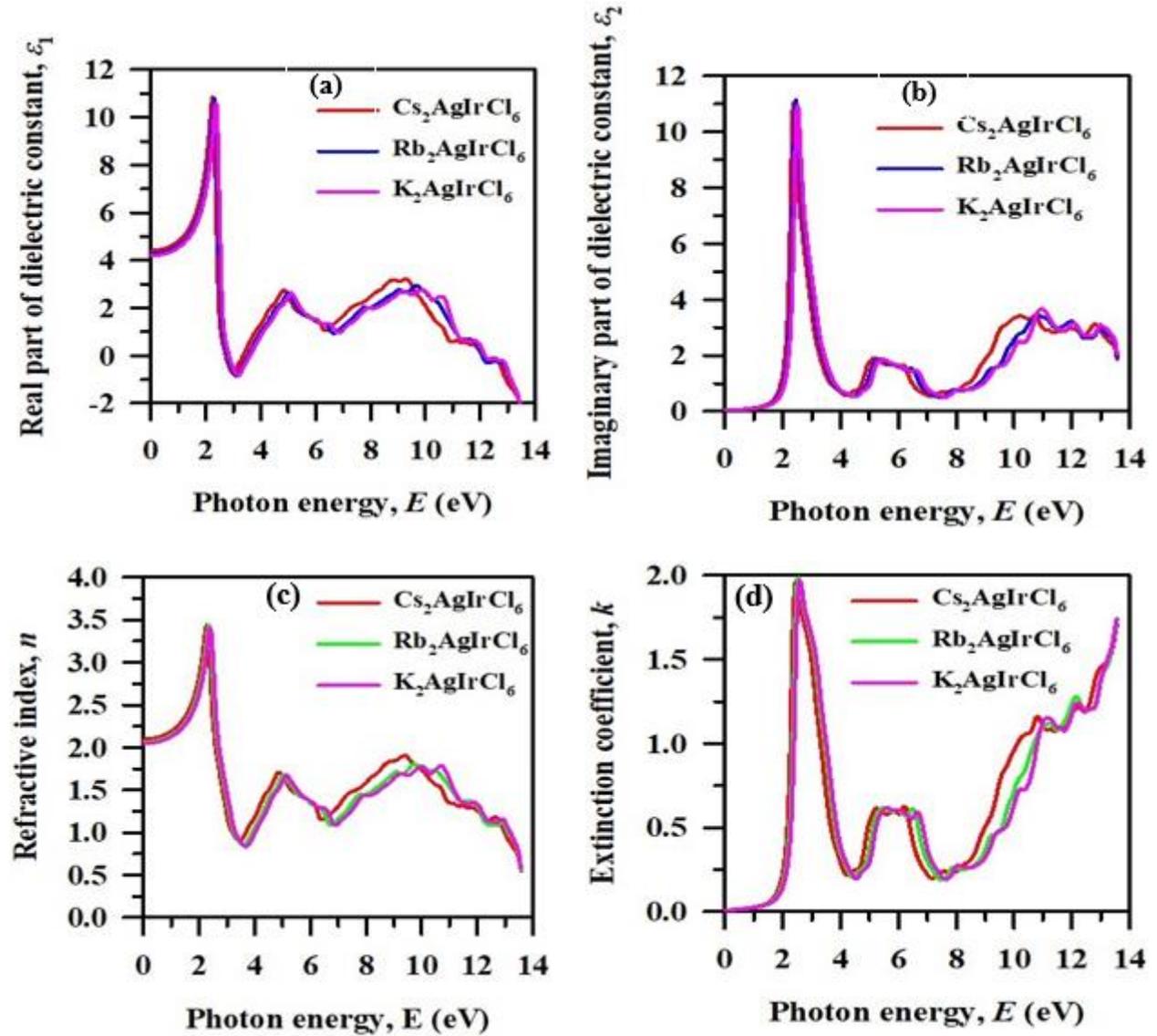

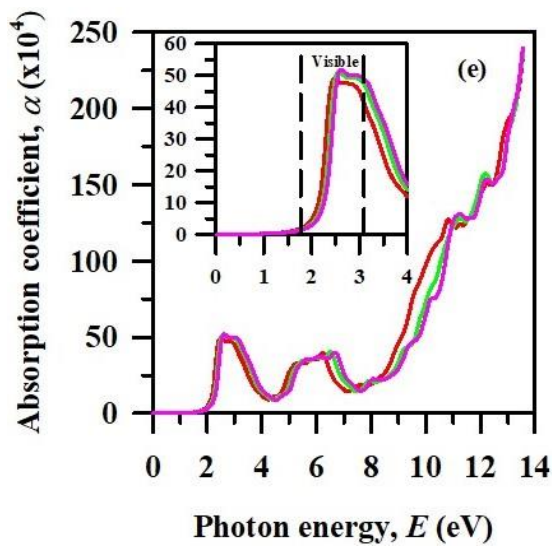
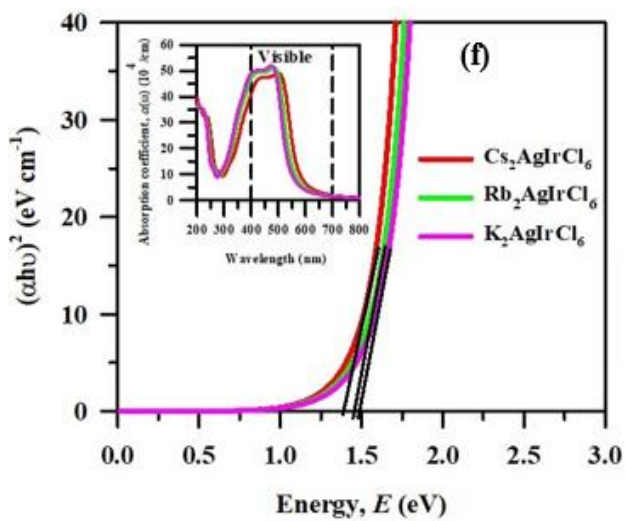
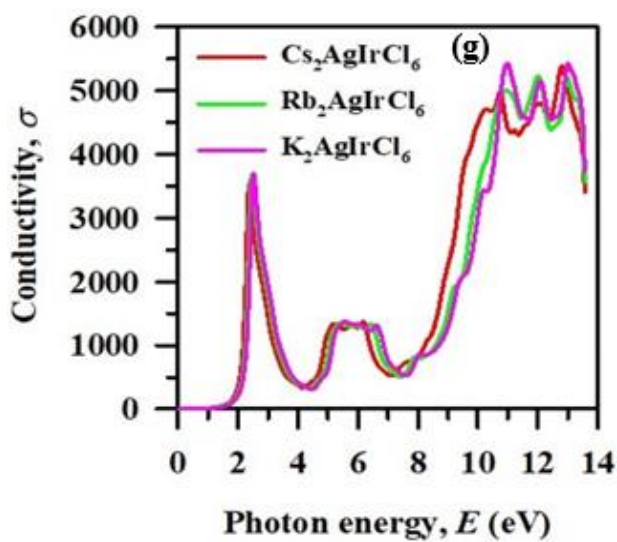
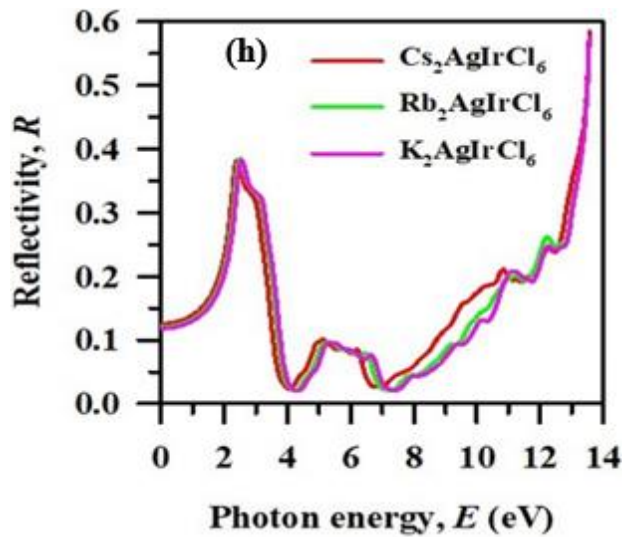
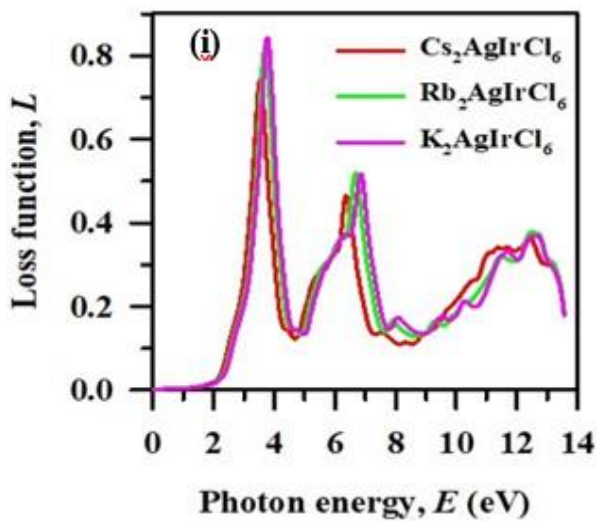

Fig. 7. Optical properties of $A_2AgIrCl_6$ (A = Cs, Rb, K) compounds.

**Table 4:** The calculated values of real part of dielectric constant $\varepsilon_1(0)$, refractive index $n(0)$, reflectivity $R(0)$, and optical band gap using Tauc plot of DP $A_2AgIrCl_6$ (A = Cs, Rb, K) compounds.

| DP | $\varepsilon_1(0)$ | $n(0)$ | $R(0)$ | Optical band gap (eV) |
|---|---|---|---|---|
| $Cs_2AgIrCl_6$ | 4.41 | 2.10 | 0.125 | 1.40 |
| $Rb_2AgIrCl_6$ | 4.27 | 2.07 | 0.121 | 1.45 |
| $K_2AgIrCl_6$ | 4.20 | 2.05 | 0.118 | 1.48 |

## 3.6 Mechanical properties

Though the study of opto-electronic properties of the titled compounds is our prime motivation, we have also studied the mechanical properties with the intention of checking the mechanical stability and their ductility (ductility is always desired for fabricating any devices), which is then further extended up to elastic moduli as a routine check. The mechanical properties of a material, including its bulk modulus, shear modulus, Young's modulus, Poisson's ratio, and Pugh's ratio, provide detailed insights into its rigidity, brittleness/ductility, and toughness characteristics. These properties collectively dictate how the material reacts to strain, significantly influencing its overall mechanical performance. The change in the A site from Cs to Rb to K in $A_2AgIrCl_6$ (A = Cs, Rb, K) induces a shift in the lattice parameter, significantly impacting the material's elastic constants, as outlined in Table 5. It is imperative to assess mechanical stability before delving into the analysis of these characteristics. Typically, $C_{ij}$ coefficients are used to provide the criteria for Born stability, which determines the mechanical stability of a lattice [83,84]. These coefficients play an important role in determining the structural integrity of a material under varied mechanical stresses. $C_{11}$, $C_{12}$, and $C_{44}$ are the three particular elastic constants ($C_{ij}$) of the cubic double perovskite structure. Determining the lattice's mechanical stability requires knowledge of these constants. For a cubic crystal, Born stability conditions are $C_{11} - C_{12} > 0$, $C_{11} + 2C_{12} > 0$, and $C_{44} > 0$, which represent the Born, spinodal, and shear criteria, respectively. To be more precise, the spinodal criterion-which is connected to the bulk modulus-needs to be positive in order to guarantee the mechanical stability of the material [85]. Table 5 shows that $A_2AgIrCl_6$ (A = Cs, Rb, K) meets all Born, spinodal, and shear criteria, demonstrating mechanical stability. Compared to other considered perovskites, the $Rb_2AgIrCl_6$ double perovskite exhibits the highest elastic $C_{ij}$ value. The ductile/brittle nature of a substance can be estimated using the Cauchy pressure (C″ = $C_{12}-C_{44}$), where a positive value signifies ductility, and a negative value indicates brittleness. Table 5 demonstrates consistently positive Cauchy pressure values for all compounds in this study. $K_2AgIrCl_6$ has the highest Cauchy pressure, while $Cs_2AgIrCl_6$ has the lowest, suggesting that $K_2AgIrCl_6$ is the most ductile and $Cs_2AgIrCl_6$ is the least ductile among these compounds. The Cauchy pressure of $A_2AgIrCl_6$ (A = Cs, Rb, K) double halide perovskite decreases as the size of the A site increases, leading to less ductile compounds.

The Shear modulus (*G*), Bulk modulus (*B*), *B/G*, Young modulus (*Y*), and Poisson's ratio (*ν*) are calculated based on elastic constants using the Voigt-Reuss-Hill approximation [86] (refer to Table 5).

$$B_V = B_R = \frac{1}{3}(C_{11} + 2C_{12}), \quad B = \frac{(B_R + B_V)}{2} \tag{15}$$

$$G_V = \frac{1}{5}(C_{11} - C_{12} + 3C_{44}), \quad G_R = \frac{5C_{44}(C_{11} - C_{12})}{4C_{44} + 3(C_{11} - C_{12})} \quad and \quad G = \frac{G_R + G_V}{2} \tag{16}$$

$$Y = \frac{9BG}{(3B + G)} \tag{17}$$

$$\nu = \frac{3B - 2G}{2(3B + G)} \tag{18}$$

The values assigned to *Y*, namely 60.70, 59.26, and 54.63 GPa for $Cs_2AgIrCl_6$, $Rb_2AgIrCl_6$, and $K_2AgIrCl_6$, respectively, serve as indicators of the respective materials' rigidity. Notably, the higher *Y* value for $Cs_2AgIrCl_6$ implies a greater stiffness when compared to $Rb_2AgIrCl_6$, and $K_2AgIrCl_6$. The observation that $Rb_2AgIrCl_6$ exhibits a larger *B* value than other two compounds further reinforce its superior resistance to deformation. Moving on to brittleness and ductility classification via the parameter *ν*, the Poisson's ratio with a threshold value of 0.26 [87], $Cs_2AgIrCl_6$, $Rb_2AgIrCl_6$, and $K_2AgIrCl_6$ are categorized as ductile materials with *ν* values of 0.27, 0.29, and 0.30, respectively. The *B/G* ratio, another crucial factor indicative of ductility or brittleness [88], supports the notion of ductility for $Cs_2AgIrCl_6$, $Rb_2AgIrCl_6$, and $K_2AgIrCl_6$, with computed values of 1.81, 2.01, and 2.12, respectively. These values align with the calculated *ν* values, collectively suggesting a susceptibility to heat shocks. The positive Cauchy pressure values, specifically 1.37 for $Cs_2AgIrCl_6$, 3.97 for $Rb_2AgIrCl_6$, and 4.78 GPa for $K_2AgIrCl_6$, further confirm the ductile nature of the studied compounds, providing additional validation for their resistance to external stresses. An important metric for material characterization is the Zener anisotropy index ($A_Z$), determined by the relation $A_Z = 2C_{44}/(C_{11} - C_{12})$ [89]. The resulting $A_Z$ values of 0.51 for $Cs_2AgIrCl_6$, 0.39 for $Rb_2AgIrCl_6$, and 0.34 for $K_2AgIrCl_6$ unequivocally classify all the studied materials as anisotropic in nature. This aligns with findings from other perovskite materials, such as $Cs_2CuIrF_6$ [33], $A_2CuSbX_6$ (A = Cs, Rb, K; X = Cl, Br, I) [78], $Cs_2AgBiX_6$ (X = Cl, Br, I) [80] which were identified as mechanically stable, hard, incompressible, and anisotropic.

Additionally, the corresponding anisotropic elastic parameters for each phase may be used to determine the characteristics of homogeneous isotropic and anisotropic polycrystals, such as $B$, $G$, $Y$, and $v$ (see Table 6). The ELATE software is utilized to calculate these attributes [90]. Additional proof of anisotropic characteristics can be seen in the 3D contour plots for $A_2AgIrCl_6$ (A = Cs, Rb, K) in Figure 8. Anisotropy is shown in $Y$ when $A$ is greater than one, meaning that none of the materials under study are spherical. A symmetric spherical shape exhibiting isotropy in linear compressibility ($\beta$) is generated when $A = 1$. Based on the $\beta$ values, it appears that the materials being studied behave in an isotropic manner. Moreover, anisotropy is shown in the ranges of maximum and least deformation for $G$ and $v$ under applied stresses. $Cs_2AgIrCl_6 <  Rb_2AgIrCl_6 < K_2AgIrCl_6$ is the order of anisotropy in elastic moduli.

**Table 5:** The evaluated elastic constants $C_{ij}$ (GPa) with Born stability criteria, Cauchy pressure, bulk modulus $B$ (GPa), Shear modulus $G$ (GPa), Young modulus $Y$ (GPa), Poisson's ratio $v$, Pugh's ratio $B/G$, and anisotropy coefficient $A_Z$ of $A_2AgIrCl_6$ (A = Cs, Rb, K) compounds.

| | Parameters | $Cs_2AgIrCl_6$ | $Rb_2AgIrCl_6$ | $K_2AgIrCl_6$ |
|---|---|---|---|---|
| **Born stability** | $C_{11}$ (GPa) | 90.86 | 99.71 | 97.74 |
| | $C_{12}$ (GPa) | 19.55 | 19.59 | 18.15 |
| | $C_{44}$ (GPa) | 18.18 | 15.62 | 13.37 |
| | $C_{11} - C_{12}$ (GPa) | 71.31 | 80.12 | 79.59 |
| | $C_{11} + 2C_{12}$ (GPa) | 129.96 | 138.89 | 134.04 |
| **Cauchy pressure, $C_P$ (GPa)** | | 1.37 | 3.97 | 4.78 |
| **Bulk modulus, $B$ (GPa)** | | 43.32 | 46.30 | 44.68 |
| **Shear modulus, $G$ (GPa)** | | 23.97 | 23.03 | 21.07 |
| **Young modulus, $Y$ (GPa)** | | 60.70 | 59.26 | 54.63 |
| **Poisson's ratio, $v$** | | 0.27 | 0.29 | 0.30 |
| **Pugh's ratio, $B/G$** | | 1.81 | 2.01 | 2.12 |
| **Zener anisotropy index, $A_Z$** | | 0.51 | 0.39 | 0.34 |

**Table 6:** Minimal and maximal values of elastic modulus and elastic anisotropy of $A_2AgIrCl_6$ (A = Cs, Rb, K) compounds.

| HDP's | Young's modulus (GPa) | | Linear compressibility (TPa$^{-1}$) | | Shear modulus (GPa) | | Poisson's ratio | |
|---|---|---|---|---|---|---|---|---|
| | $Y_{min}$ | $Y_{max}$ | $\beta_{min}$ | $\beta_{max}$ | $G_{min}$ | $G_{max}$ | $\nu_{min}$ | $\nu_{max}$ |
| $Cs_2AgIrCl_6$ | 47.85 | 83.94 | 7.69 | 7.69 | 18.18 | 35.65 | 0.11 | 0.47 |
| $Rb_2AgIrCl_6$ | 41.86 | 84.81 | 7.64 | 7.64 | 15.62 | 36.06 | 0.10 | 0.53 |
| $K_2AgIrCl_6$ | 36.47 | 92.05 | 7.46 | 7.46 | 13.37 | 39.79 | 0.07 | 0.61 |
| **Elastic anisotropy $A_x$** | | | | | | | | |
| | $A_Y$ | | $A_\beta$ | | $A_G$ | | $A_\nu$ | |
| $Cs_2AgIrCl_6$ | 1.75 | | 1.00 | | 1.96 | | 4.19 | |
| $Rb_2AgIrCl_6$ | 2.03 | | 1.00 | | 2.31 | | 5.37 | |
| $K_2AgIrCl_6$ | 2.52 | | 1.00 | | 2.98 | | 8.30 | |

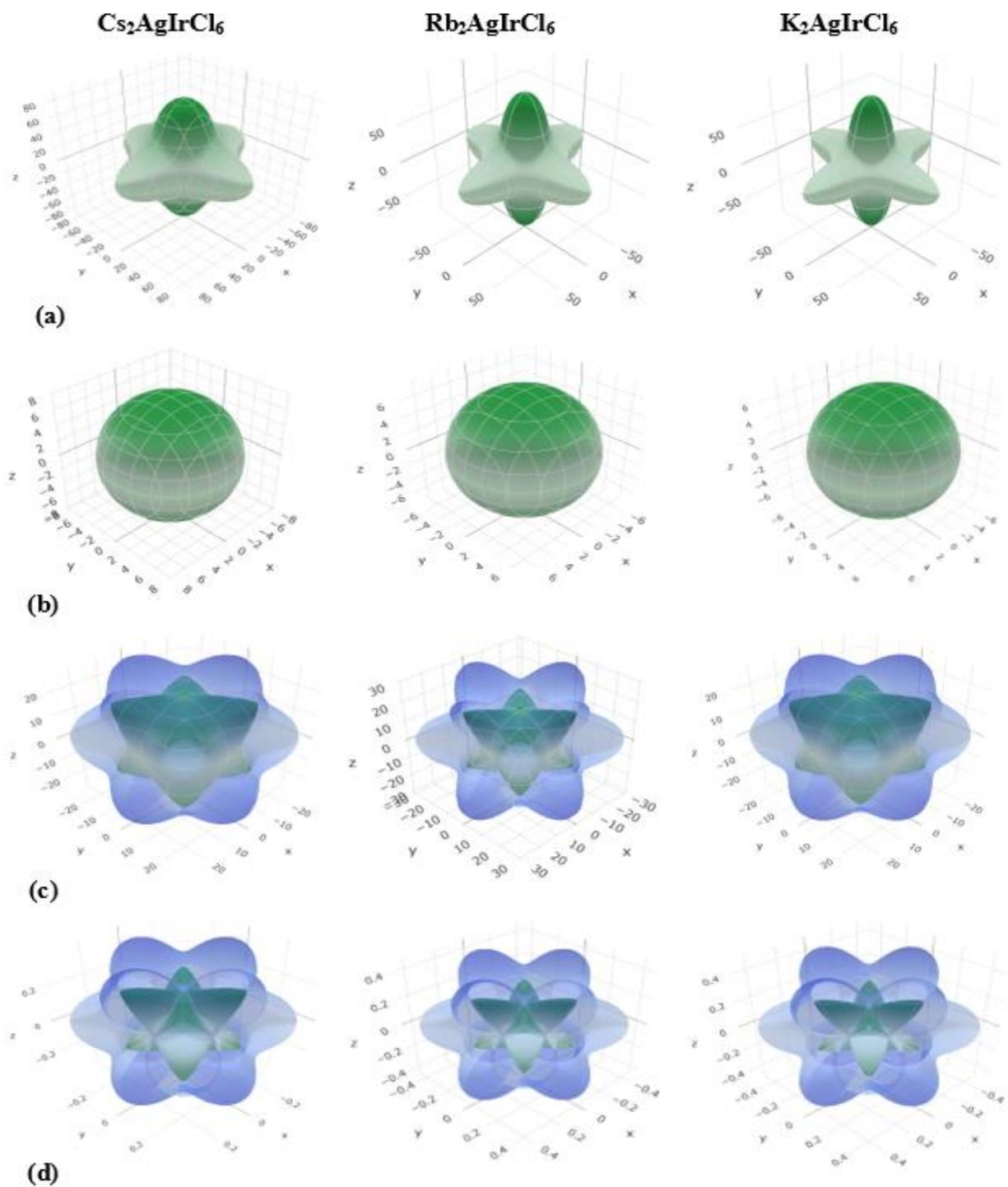

Fig. 8. 3D directional dependences of a) Young's modulus, b) linear compressibility, c) shear modulus, and d) Poisson's ratio for the compounds $Cs_2AgIrCl_6$, $Rb_2AgIrCl_6$, and $K_2AgIrCl_6$.

## 3.7  Thermal properties

At present, the world faces significant challenges, notably global warming, and the dwindling reserves of renewable energy sources. These issues are exacerbated by emissions from fossil fuel combustion and the limited nature of traditional energy supplies. One promising solution involves the conversion of solar energy into usable forms through photovoltaic cells [91–93]. One important thermodynamic parameter that is necessary to comprehend the characteristics and heat capacity of solids at various temperatures is the Debye temperature ($\theta_D$). It provides information on how material properties change as a result of temperature changes. To compute $\theta_D$, a number of theoretical models have been established; these models usually use formulas that take into account the average sound velocity ($v_m$) [94]. This parameter plays a crucial role in studying and predicting the thermal behavior and stability of materials.

$$\theta_D = \frac{\hbar}{k_B} \left[\frac{3n}{4\pi V}\right]^{\frac{1}{3}} \times v_m \tag{19}$$

In the equation mentioned, $\hbar$ denotes the normalized Planck's constant, $k_B$ signifies Boltzmann's constant, $n$ stands for the number of atoms, $V$ indicates the volume, and $v_m$ refers to the average velocity of sound. Additionally, $v_m$ can be determined using the following formula [94]:

$$v_m = \frac{1}{3}\left[\frac{2}{v_t^3} + \frac{1}{v_l^3}\right]^{-1/3} \tag{20}$$

In the above expression, $v_t$ represents the transverse sound velocity, while $v_l$ denotes the longitudinal sound velocity. Both $v_t$ and $v_l$ can be computed using the given equation [94]:

$$v_t = \left[\frac{G}{\rho}\right]^{1/2} \tag{21}$$

$$v_l = \left[\frac{3B + 4G}{3\rho}\right]^{1/2} \tag{22}$$

The Debye temperatures have been found to be 246.24 K, 256.63 K, and 263.82 K for $Cs_2AgIrCl_6$, $Rb_2AgIrCl_6$, and $K_2AgIrCl_6$, respectively.

In the manufacturing of solar cells, materials are subjected to high temperatures, such as during the growth of crystalline silicon or the formation of metal contacts. If the melting temperature of a material is too low, it can lead to deformation, cracking, or melting, resulting in defects and reduced performance. Conversely, if the melting temperature is too high, the process becomes more complex and costlier. The elastic constant $C_{11}$ is used to assess the melting temperatures of these materials and can be calculated using the following expression [95]:

$$T_m = (553 + 5.91 C_{11}) K \tag{23}$$

The calculated melting temperatures imply that these double perovskite materials are feasible to synthesize at ambient conditions. Among them, Rb$_2$AgIrCl$_6$ exhibits a higher melting point compared to Cs$_2$AgIrCl$_6$ and K$_2$AgIrCl$_6$. The values of thermodynamic properties, as determined from the elastic constants, are summarized in Table 7.

**Table 7:** The computed density ($\rho$), longitudinal, transverse, and average sound velocities ($v_l$, $v_t$ and $v_m$, respectively), Debye temperature ($\theta_D$), melting temperature ($T_m$) of A$_2$AgIrCl$_6$ (A = Cs, Rb, K) compounds.

| Parameters | Cs$_2$AgIrCl$_6$ | Rb$_2$AgIrCl$_6$ | K$_2$AgIrCl$_6$ |
|---|---|---|---|
| $\rho \times 10^3$ (Kg/m$^3$) | 4.89 | 4.42 | 3.89 |
| $v_l$ (m/s) | 3930.01 | 4164.74 | 4327.21 |
| $v_t$ (m/s) | 2217.58 | 2277.51 | 2328.53 |
| $v_m$ (m/s) | 2466.74 | 2539.68 | 2599.67 |
| $\theta_D$ (K) | 246.24 | 256.63 | 263.82 |
| $T_m$ (K) | 1089.97 | 1142.31 | 1130.65 |

We assessed the stability of the sample by analyzing its phonon modes at various temperatures, which allowed us to calculate key thermodynamic values like enthalpy, free energy, and entropy. These values are influenced by the frequency of phonon vibrations, which are derived by [96]:

$$H(T) = E_{tot} + \frac{1}{2} \int g(\omega) \hbar \omega d\omega + \int \frac{\hbar \omega}{e^{\frac{\hbar \omega}{K_B T}} - 1} g(\omega) d\omega \tag{24}$$

$$F(T) = E_{tot} + \frac{1}{2} \int g(\omega) \hbar \omega d\omega + K_B T \int g(\omega) \ln\left(1 - e^{\frac{\hbar \omega}{K_B T}}\right) d\omega \tag{25}$$

$$S(T) = K_B \left[ \int \frac{\frac{\hbar \omega}{K_B T}}{e^{\frac{\hbar \omega}{K_B T}} - 1} g(\omega) d\omega - \int g(\omega) \ln\left(1 - e^{\frac{\hbar \omega}{K_B T}}\right) d\omega \right] \tag{26}$$

In this context, $g(\omega)$ represents the phonon density of states, $k_B$ stands for the Boltzmann's constant, $E_{tot}$ denotes the minimum total energy, and $\hbar$ refers to the reduced Planck constant. Figure 9 shows how the predicted thermodynamic parameters of the A$_2$AgIrCl$_6$ (A = Cs, Rb, K) with temperature owing to phonon states. These variables include enthalpy, free energy, entropy, heat capacity, and Debye temperature. The data shows that while free energy falls with temperature, enthalpy, entropy, and heat capacity all rise.

Additionally, we note a more significant enhancement in the heat capacity of the $A_2AgIrCl_6$ (A = Cs, Rb, K) compared to changes in enthalpy and entropy, suggesting its heightened sensitivity to temperature variations. According to the data in Fig. 9(a), the enthalpy, entropy, and free energy values at room temperature are determined as 0.56 (0.56, 0.47) eV, 1.33 (1.36, 1.13) eV, and 0.77 (0.80, 0.66) eV, respectively, for $Cs_2AgIrCl_6$ ($Rb_2AgIrCl_6$, $K_2AgIrCl_6$). The computed enthalpy values at 1000 K are 2.35 eV, 2.34 eV, and 2.07 eV for $Cs_2AgIrCl_6$, $Rb_2AgIrCl_6$, and $K_2AgIrCl_6$, respectively. The heat capacity gradually rises with temperature and eventually stabilizes at 59.45 (58.85, 53.03) cal/cell.K for $Cs_2AgIrCl_6$ ($Rb_2AgIrCl_6$, $K_2AgIrCl_6$) shown in Fig. 9(b), a phenomenon known as the Dulong-Petit limit [97]. The variation of Debye temperature as a function of temperature is shown in Fig. 9(c). For the compounds $Cs_2AgIrCl_6$, $Rb_2AgIrCl_6$, and $K_2AgIrCl_6$, the estimated Debye temperature at 0 K is about 116.96 K, 85.87 K, and 122.04 K, respectively. Under constant pressure, the Debye temperature rises with temperature. These thermodynamic characteristics provide insights for comparing with experimental results to predict phase stability.

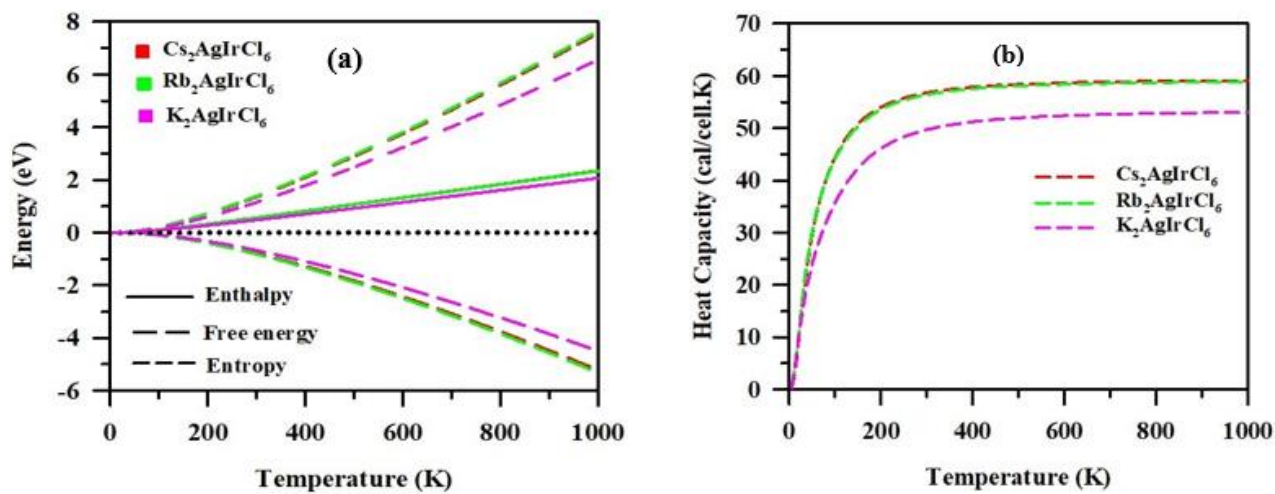

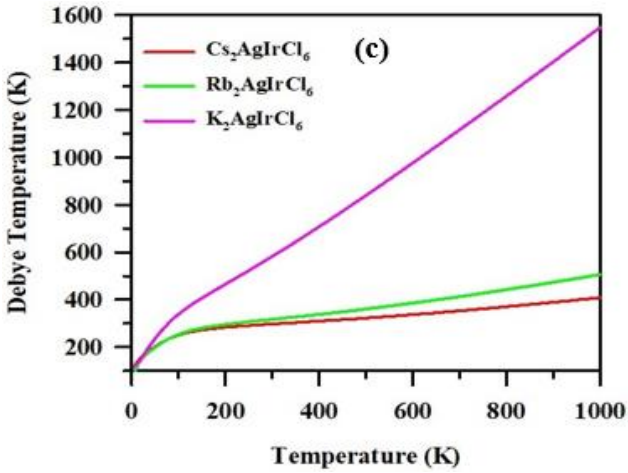

Fig. 9. Thermodynamic properties of $A_2AgIrCl_6$ (A = Cs, Rb, K) perovskite.

## 4. Conclusions

In this study, three new double perovskite halides $A_2AgIrCl_6$ (A = Cs, Rb, K) have been explored for the first time, and the structural, electronic, optical, mechanical, and thermo-physical properties are studied. The stable cubic structure of the herein-predicted compounds has been confirmed under $Fm\bar{3}m$ symmetry through energy optimization and tolerance factor calculations. The electronic properties of these materials show semiconducting features; at X-X symmetry k-points, a direct band gap is detected. The computed energy band gaps for $Cs_2AgIrCl_6$, $Rb_2AgIrCl_6$, and $K_2AgIrCl_6$ are 1.43 eV, 1.50 eV, and 1.55 eV, respectively, which makes them well-suited for solar cell materials. From the band structure and effective mass calculations, it has been confirmed that all the compounds show a p-type nature. The effective mass of electrons is always smaller than the effective mass of holes for all the studied compounds, aiding efficient carrier transport properties. In addition, the study of the optical properties revealed their strong visible light absorption characteristics (up to $10^5$) and very low reflectivity, which suggest their promise for use as an absorber layer in solar cell technology. Mechanical properties, including Pugh's ratio, Cauchy's pressure, and Poisson's ratio, affirmed their ductility. The elastic constants-derived thermodynamic analysis revealed that the Debye temperatures of the materials exhibit the trend: $K_2AgIrCl_6 > Rb_2AgIrCl_6 > Cs_2AgIrCl_6$. Thus, the newly predicted double perovskites $A_2AgIrCl_6$ (A = Cs, Rb, K) with excellent opto-electronic properties are potential candidates for photovoltaic applications and this study presents valuable insights that can serve as a guide in designing and predicting new materials.

**Acknowledgments**

This work was carried out with the aid of a grant (grant number: 21-378 RG/PHYS/AS_G - FR3240319526) from UNESCO-TWAS and the Swedish International Development Co-operation Agency (SIDA). The views expressed herein do not necessarily represent those of UNESCO-TWAS, SIDA or its Board of Governors.

**CRediT Author contributions**

**M. A. Rayhan:** Writing – original draft, Methodology, Conceptualization, Formal analysis, Data calculations, Validation. **M.M. Hossain:** Writing – review & editing, Validation. **M.M. Uddin:** Writing – review & editing, Validation. **S. H. Naqib:** Writing – review & editing, Validation. **M.A. Ali:** Conceptualization, Formal analysis, Validation, Writing – review & editing, Supervision, Software.

*Chem. Phys.* **18**, 23872–23878 (2016).